\documentclass[aps,prl,twocolumn,preprintnumbers,nofootinbib]{revtex4}
\usepackage{graphicx}
\usepackage{amssymb}
\usepackage{amsmath,mathrsfs,verbatim}
\usepackage{times}
\usepackage{latexsym}
\def\beq{\begin{equation}}
\def\eeq{\end{equation}}
\def\bey{\begin{eqnarray}}
\def\eey{\end{eqnarray}}

\def\lsim{\mathrel{\raise.3ex\hbox{$<$\kern-.75em\lower1ex\hbox{$\sim$}}}}
\def\gsim{\mathrel{\raise.3ex\hbox{$>$\kern-.75em\lower1ex\hbox{$\sim$}}}}

\newcommand{\be}{\begin{equation}}
\newcommand{\ee}{\end{equation}}

\newcommand{\mx}{m_X}
\newcommand{\ax}{\alpha_X}
\newcommand{\mphi}{m_\phi}

\newcommand{\cmg}{{\rm cm}^2/{\rm g}}

\begin{document}

\preprint{MCTP/12-27}

\title{Resonant Dark Forces and Small Scale Structure}

\author{Sean Tulin$^1$, Hai-Bo Yu$^1$, and Kathryn M. Zurek$^{1,2}$}
\affiliation{$^1$ Department of Physics, University of Michigan, Ann Arbor, MI 48109  \\ $^2$ School of Natural Sciences, Institute for Advanced Study, Princeton, NJ 08540}

\date{\today}

\begin{abstract}

A dark force can impact the cosmological history of dark matter (DM), both explaining observed cores in dwarf galaxies and setting the DM relic density through annihilation to dark force bosons.  For GeV -- TeV DM mass, DM self-scattering in dwarf galaxy halos exhibits quantum mechanical resonances, analogous to a Sommerfeld enhancement for annihilation.  We show that a simple model of DM with a dark force can accommodate all astrophysical bounds on self-interactions in halos and explain the observed relic density, through a single coupling constant.

\end{abstract}

\maketitle

{\em I. Introduction:}  The paradigm of cold, collisionless dark matter (DM) has been extraordinarily successful in explaining astrophysical observations of structure, from the recombination epoch to the present large scale structure of the Universe.  
Although all evidence for DM is from its gravitational influence, it is expected that DM possesses some type of interactions beyond gravity.  Non-gravitational interactions are required to produce DM particles in the early Universe, and ultimately determine the DM density observed today. 

Despite its great success, it is unclear whether cold, collisionless DM can successfully account for the small scale structure of the Universe, which may indicate other interactions besides gravity play a role in structure formation. Precision observations of dwarf galaxies by THINGS show DM mass distributions with flat cores, compared to steep cusps predicted by collisionless DM simulations~\cite{Oh:2010ea}.  The gravitational effect of massive baryonic outflows from supernovae can potentially flatten central DM cusps~\cite{Oh:2010mc}, but it is unknown whether this effect can explain the observed cores in other less luminous (more DM-dominated) dwarf galaxies~\cite{Goerdt:2006rw}.  Another discrepancy is the apparent underabundance of Milky Way (MW) satellite dwarf galaxies, compared to predictions from collisionless DM simulations~\cite{Klypin:1999uc}.  The missing low mass satellites may simply be fainter than expected if energy injection from astrophysical processes strips away interstellar gas and suppresses star formation~\cite{Koposov:2009ru}.  However, this mechanism cannot explain the apparent absence of the most massive subhalos predicted by simulations~\cite{Sawala:2010zw} which are ``too big to fail'' in star formation and are too dense to host any observed MW satellite, according to their predicted stellar circular velocities~\cite{BoylanKolchin:2011de}.

These small scale structure anomalies can be explained if DM, denoted $X$, is self-interacting~\cite{Spergel:1999mh}. 
Recent N-body simulations have shown that a DM self-interaction cross section per unit mass $\sigma_T/m_X \sim 0.1-10\, \cmg$ can flatten the central density within dwarf galaxies and subhalos to solve the core/cusp problem~\cite{Vogelsberger:2012ku,Rocha:2012jg}.  Moreover, the most massive subhalos can be reconciled with the observed MW satellites since stellar circular velocities are reduced in their central cores~\cite{Vogelsberger:2012ku,Rocha:2012jg}. 
At the same time, a variety of constraints from larger scales ({\it e.g.}, halo shapes of elliptical galaxy and cluster halos~\cite{MiraldaEscude:2000qt}, the Bullet cluster~\cite{Randall:2007ph}, subhalo evaporation~\cite{Gnedin:2000ea}) have suggested that $\sigma_T/m_X$ must be smaller on these scales, motivating a velocity-dependent force~\cite{Feng:2009hw,Loeb:2010gj} that gives $\sigma_T/m_X \sim 10 \, \cmg$ on dwarf scales~\cite{Vogelsberger:2012ku}, but is suppressed on larger scales.  However, simulations with a constant cross section have shown that the aforementioned constraints are in fact much weaker than previously thought, and a constant $\sigma_T/m_X \sim 0.1 \, \cmg$ is sufficient to solve small scale structure anomalies while evading other bounds~\cite{Rocha:2012jg}.

Given these results, it is important to explore the particle physics nature of DM self-interactions. For typical weakly-interacting DM models, self-scattering has a weak-scale cross section $\sigma_T \sim 10^{-36}\,{\rm cm}^2$, far too small to play a role in galactic dynamics.  An MeV-scale dark force mediator (denoted $\phi$) is needed to give a much larger scattering cross section, 
\mbox{$\sigma_T \sim 1 \, {\rm cm}^2 \, (m_X/{\rm g}) \approx 2 \!\times\! 10^{-24} \, {\rm cm}^2\, (m_X/{\rm GeV})$}, required to leave observable signatures on DM halos~\cite{Feng:2009hw,Buckley:2009in,Loeb:2010gj,Lin:2011gj,Aarssen:2012fx}.  A perturbative calculation for $\sigma_T$ from $\phi$ exchange gives $\sigma_T \approx 4\pi \alpha_X^2 m_X^2/m_\phi^4$ at small velocity, where $\alpha_X$ is the ``dark fine structure constant,''  so that
\be
\sigma_T \approx 5 \times 10^{-23} \, {\rm cm}^2 \,  \left(\frac{\alpha_X}{0.01}\right)^2 \left(\frac{m_X}{{\rm 10\,  GeV}} \right)^2 \left(\frac{{\rm 10 \,MeV}}{m_\phi} \right)^4  \label{rough}
\ee
in the desired range.  However, this calculation breaks down for $m_\phi \lesssim \alpha_X m_X$, and nonperturbative effects become important.  These effects have not been studied in general and yet are crucial for connecting dark forces to small scale structure.  In particular, DM self-scattering exhibits quantum mechanical resonances, analogous to resonant Sommerfeld enhancement for annihilation, as we show below.

In this {\it Letter}, we present a simple model where a dark force can simultaneously set the DM abundance and solve small scale structure anomalies. In the early Universe, $X \bar X \to \phi \phi$ provides an efficient annihilation channel for obtaining the relic density during freeze-out.  This same coupling resolves structure problems through scattering on small scales while remaining consistent with bounds on MW and cluster scales.  We consider both symmetric and asymmetric DM models which involve attractive and repulsive DM self-interactions. In calculating the scattering cross section, we take a numerical approach and cover the full parameter space including the nonperturbative quantum mechanical regime which has not been explored before.  We show that resonances can arise for a wide range of DM mass, $m_X \sim {\rm GeV - TeV}$.  Furthermore, our numerical calculation confirms analytical formulae widely used in literature for computing $\sigma_T$ in the classical and Born limits.

{\em II. DM Annihilation and Elastic Scattering:} 
We consider a Dirac fermion DM particle $X$, coupled to a dark force vector boson $\phi$ with mass $m_\phi$ via
\be
\mathscr{L}_{\rm int} = g_X \bar X \gamma^\mu X \phi_\mu,
\ee
where $g_X$ is the coupling constant.  We assume that $X$ is weakly coupled to the SM ({\it e.g.}, through kinetic mixing of $\phi$ with $U(1)_Y$ hypercharge) so that $X$ thermalizes with the visible sector in the early Universe~\cite{Feng:2010zp}.  

DM freeze-out is governed by the velocity-weighted annihilation cross section for $X \bar{X} \to \phi \phi$, given by $\langle \sigma v \rangle_{\rm an}\approx{\pi \alpha_X^2}/{m_X^2} \,$ where $\alpha_X \equiv g_X^2/(4\pi)$.  For {\it symmetric} DM, where DM consists of equal densities of $X$ and $\bar{X}$, we require $\langle \sigma v \rangle_{\rm an} \approx 6 \times 10^{-26} \, {\rm cm}^3/{\rm s}$ to obtain the observed relic density.  For {\it asymmetric} DM, the present DM density is determined by a primordial asymmetry between $X$ and $\bar X$, in analogy to the baryon asymmetry.  
In this case, we require larger $\langle \sigma v\rangle_{\rm an}$ to deplete the symmetric $X,\bar X$ density, leaving behind only the residual asymmetric $X$ density as DM.  Thus, we have $\alpha_X \gtrsim 4 \times 10^{-5} \: (m_X/{\rm GeV})$, with the lower bound saturated for symmetric DM.  Asymmetric DM allows for a broader region of parameter space, since annihilation $X \bar X \to \phi \phi$ sufficient to set the relic density only places a lower bound on $\alpha_X$, rather than fixing it to a particular value as a function of $m_X$. 

In our model, the same dark force carrier $\phi$ also mediates DM self-interactions. Here, the relevant quantity is the  scattering cross section weighted by the momentum transfer, $\sigma_T = \int d\Omega \, (1-\cos\theta) \, d\sigma/d\Omega$,
where $d\sigma/d\Omega$ is the usual differential cross section. The nonrelativistic interaction between two DM particles mediated by $\phi$ is described by a Yukawa potential 
\be
V(r) = \pm \frac{\alpha_X}{r} e^{-m_\phi r}.
\ee
Since $\phi$ is a vector, $X X \to X X$ scattering is repulsive ($+$), while $X \bar X \to X \bar X$ is attractive ($-$). For symmetric DM, {\it both} attractive ($X$-$\bar X$) and repulsive ($X$-$X$ or $\bar X$-$\bar X$) interactions are present; for asymmetric DM, where DM consists of only $X$ after the freeze-out, self-interactions are {\it only} repulsive.  

Since both scattering and annihilation occur through a common interaction, the cross sections are related.  When $\phi$ is massless, the scattering cross section scales roughly as $\sigma_T \sim \langle \sigma v \rangle_{\rm an}/ v^{4}$.  If this relation holds to dwarf scales ($v\sim10~{\rm km/s}$), the transfer cross section is $\sigma_T/\mx\sim10^{3}~{\rm cm^2/g}~({{\rm TeV}/\mx})$, which is too large compared to that preferred by the simulation results~\cite{Vogelsberger:2012ku,Rocha:2012jg} unless the DM mass is larger than $100~{\rm TeV}$. Therefore, a nonzero $m_\phi$ is essential, softening the velocity-dependence of $\sigma_T$ at small $v$ due to the finite range of the dark force.

\begin{figure}
\includegraphics[scale=0.7]{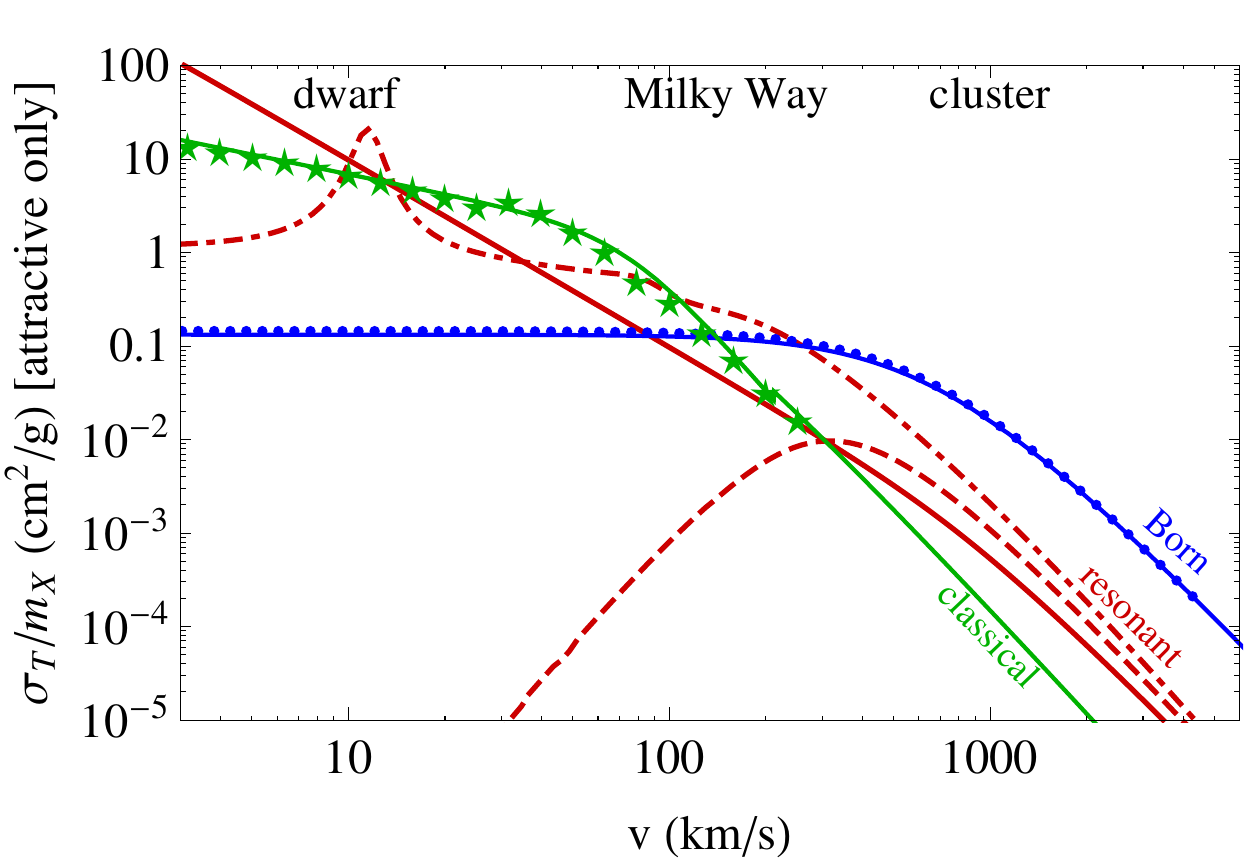}
\caption{Velocity-dependence of $\sigma_T$ for sample parameters within different regimes.  Blue line shows Born formula \eqref{born}, in agreement with numerical results (blue dots), for $m_X = 4$ GeV, $m_\phi = 7.2$ MeV, $\alpha_X = 1.8 \times 10^{-4}$.  Green line shows classical formula \eqref{plasma}, in agreement with numerical results (stars), for $m_X = 2$ TeV, $m_\phi = 1$ MeV, $\alpha_X=0.05$.  Red lines show $\sigma_T$ in the resonant regime for $m_X = 100$ GeV, $\alpha_X = 3.4\times 10^{-3}$, illustrating $s$-wave resonance (solid, $m_\phi = 205$ MeV), $p$-wave resonance (dot-dashed, $m_\phi = 20$ MeV), and $s$-wave antiresonance (dashed, $m_\phi=77$ MeV).}
\label{sigmaVv}
\end{figure}

The calculation of $\sigma_T$ for a Yukawa potential with $m_\phi\ne 0$ is non-trivial.  We collect analytical results, where applicable, in the appendix.  
Within the Born approximation (valid for $\ax\mx/\mphi\ll1$), $\sigma_T$ can be computed perturbatively.  Outside the Born regime, multiple $\phi$ scatterings lead to a nonperturbative modification of the DM two-body wavefunction, and an analytical approximation has been obtained only within the classical limit ($m_X v/m_\phi \gg 1$).  However, outside the Born and classical regimes, no analytic description is possible, and one must compute $\sigma_T$ by solving the the Schr\"{o}dinger equation numerically using a partial wave analysis~\cite{Buckley:2009in,inprep}.  Within this ``resonant'' regime, $\sigma_T$ exhibits a rich structure of quantum mechanical resonances (for the attractive potential case).\footnote{Ref.~\cite{Buckley:2009in} previously studied this effect for limited parameter choices motivated by cosmic ray excesses; here, we have adopted a more efficient numerical procedure (described in a forthcoming publication~\cite{inprep}) allowing us to explore the full parameter range in detail.} Computing $\sigma_T$ within this regime is crucial for understanding for what parameters a dark force can explain simultaneously small scale structure problems and the DM relic density.

\begin{figure*}[t!]
\includegraphics[scale=0.65]{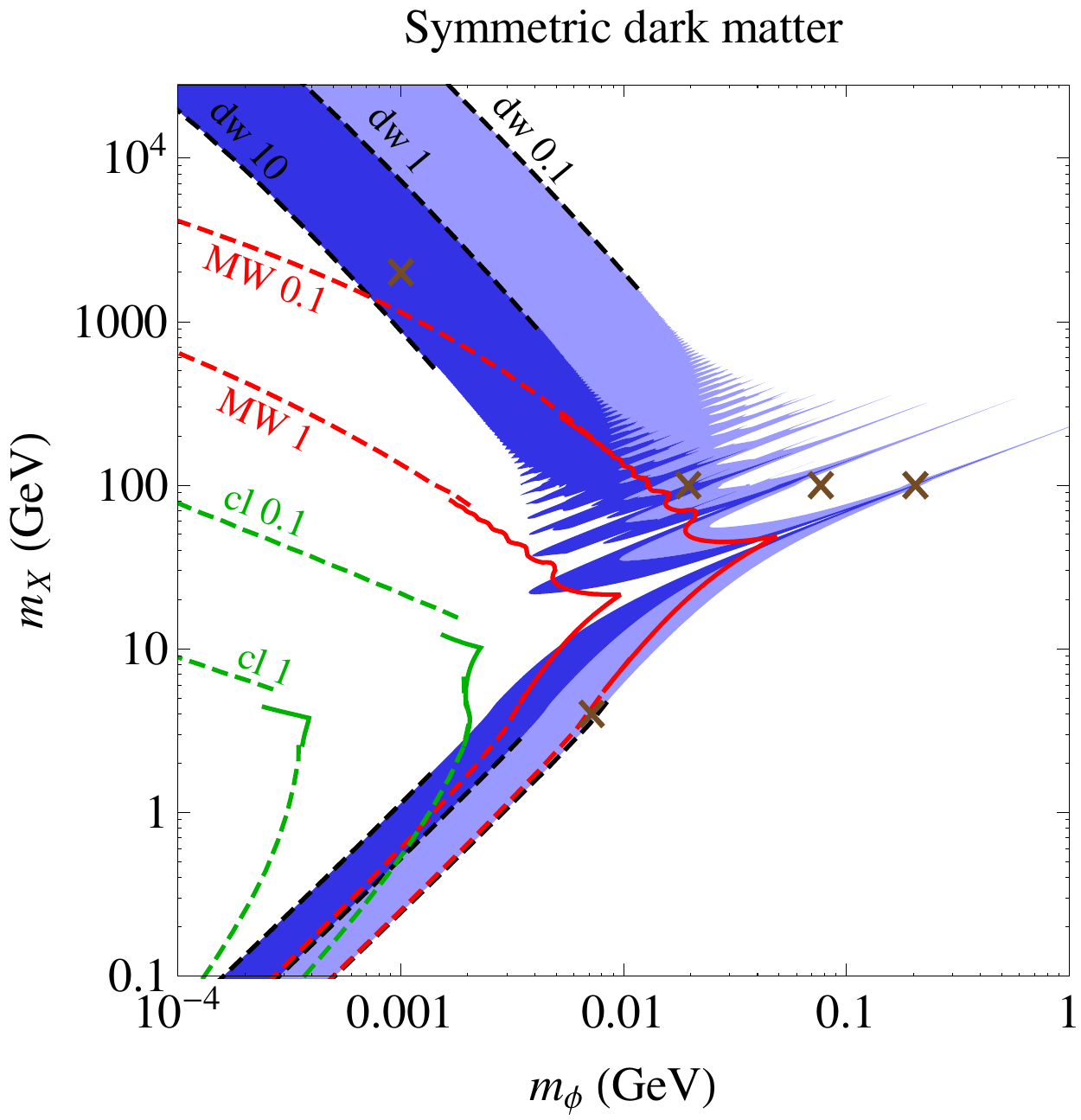}~~~
\includegraphics[scale=0.65]{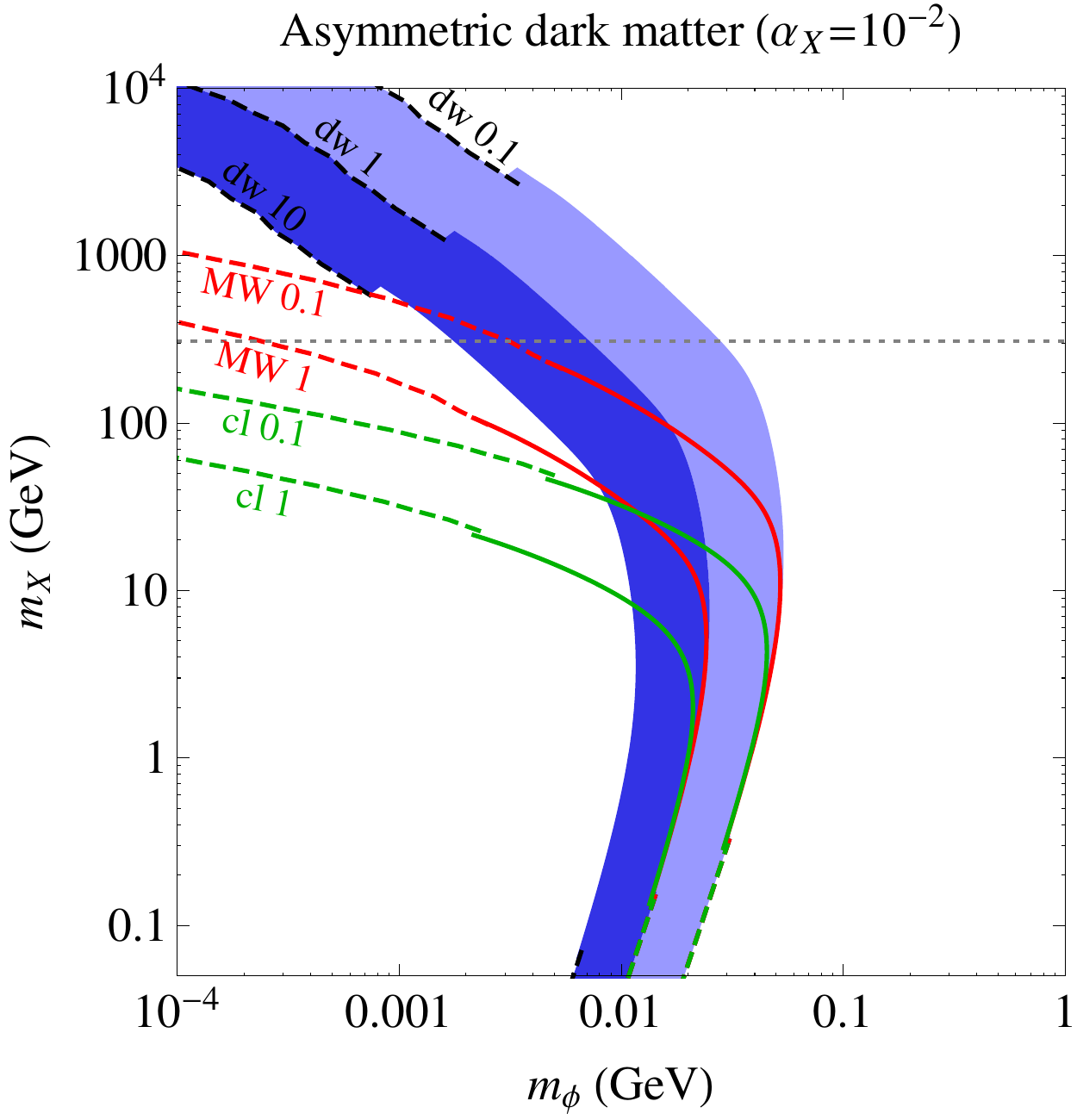}
\caption{Symmetric (left) and asymmetric (right) DM parameter space in $m_X$-$m_\phi$ plane.  Blue regions show where DM self-scattering solves dwarf-scale structure anomalies, while red (green) lines show bounds from Milky Way (cluster) scales.  Numerical values indicate $\langle \sigma_T \rangle/m_X$ in $\cmg$ on dwarf (``dw''), Milky Way (``MW''), and cluster (``cl'') scales.  For symmetric DM, $\alpha_X$ is fixed to obtain the observed relic density; for asymmetric DM, $\alpha_X = 10^{-2}$ is fixed to deplete $X,\bar X$ density for $m_X \lesssim 300$ GeV (dotted line).  Dashed lines show extrapolation using analytic formulae, while ``x'' marks parameter points utilized in Fig.~1.}
\label{ResonantPlots}
\end{figure*}

To illustrate the different regimes and behaviors of DM self-scattering, Fig.~\ref{sigmaVv} shows $\sigma_T/\mx$ as a function of $v$ for an attractive potential, for several parameter choices.  The blue (green) line shows the analytic result for $\sigma_T$ for a parameter point within the Born (classical) regime (see appendix); these formulae are in excellent agreement with our numerical results, shown by the blue dots (green stars).  The red lines correspond to three parameter points within the resonant regime.  The solid red line shows an $s$-wave resonance, with $\sigma_T$ growing as $v^{-2}$ at small velocity. The dot-dashed line shows a $p$-wave resonance, where $\sigma_T$ shows a resonant peak at finite $v$.  These two cases illustrate how $\sigma_T$ may be enhanced at dwarf scales due to resonances.  The dashed line shows an example with an {\it antiresonance} (the Ramsauer-Townsend effect), which can suppress $\sigma_T$ at small $v$.  All of these parameters have been chosen to give the correct DM relic density and $\sigma_T/m_X \sim 0.1 - 10 \, \cmg$ to solve structure problems on dwarf scales (except for the antiresonance case).

{\em III. Results:} We show the complete parameter space where a dark force can account for DM small scale structure and relic density.  
For scattering, to compare with astrophysical bounds, we consider the velocity-averaged cross section $\langle \sigma_T \rangle =  \int d^3 v \, \sigma_T \, e^{-\frac{1}{2} v^2/v_0^2 } /(2\pi v_0^2)^{3/2}$, where $v_0$ is the most probable velocity for a DM particle.  Fig.~\ref{ResonantPlots} shows contour plots of $\langle \sigma_T \rangle$ for two cases, symmetric and asymmetric DM, in the $m_X$-$m_\phi$ parameter space.  

For symmetric DM (Fig.~\ref{ResonantPlots}, {\it left}), we take the average of attractive and repulsive cross sections, $\sigma_T = (\sigma_T^{\rm att} + \sigma_T^{\rm rep})/2$, with $\alpha_X$ chosen to reproduce the observed DM relic density at each point.\footnote{We compute the relic density by solving numerically the Boltzmann equations for DM freeze-out, accounting for a possible Sommerfeld enhancement in $\langle \sigma v \rangle$.  We assume $X$ kinetically decouples at a temperature $0.5~{\rm MeV}$, {\it e.g.}, if $X$ were weakly coupled to electrons~\cite{Feng:2010zp}.}  
The blue contour regions show $\langle \sigma_T\rangle/\mx$ on dwarf scales ($v_0 = 10$ km/s) in the ranges $0.1 - 1 \, \cmg$ (light) and $1 - 10 \, \cmg$ (dark) to solve small scale structure problems.  The lower range is prefered for a constant cross section; Ref.~\cite{Rocha:2012jg} found $0.1 \, \cmg$ matched small scale structure observations, while $1 \, \cmg$ caused too low central densities in dwarf spheroidals.  Simulations with a $v$-dependent classical (attractive-only) force prefered the upper range (or larger)~\cite{Vogelsberger:2012ku}.  The red (green) contours show $\langle \sigma_T\rangle/\mx = 0.1$ and $1 \, \cmg$ on MW (cluster) scales with $v_0 = 200$ ($1000$) km/s, showing the approximate upper limits from observations.  Ref.~\cite{Rocha:2012jg} found that $1 \, \cmg$ produced a too-small central DM density in galaxy clusters and is only marginally consistent with MW-scale halo shape ellipticity constraints, while $0.1 \, \cmg$ is consistent with these constraints~\cite{Rocha:2012jg}.
In the resonant regime, we have computed $\sigma_T$ numerically.  This region shows a pattern of resonances for $m_X \sim 10$ GeV -- TeV, where $\sigma_T^{\rm att}$ is enhanced, allowing for larger $m_X$ for fixed $\langle \sigma_T \rangle/m_X$.  The dashed lines indicate where we use analytic formulae to extrapolate our results into the Born ($m_X \ll m_\phi/\alpha_X$) and classical ($m_X \gg m_\phi/v$) regimes. Our numerical calculation maps smoothly into these regions, again confirming our agreement with the analytic formulae.\footnote{The small discrepancy on cluster scales is because $\langle \sigma_T \rangle$ at these parameters is dominated by phase space with $v \ll v_0$, where the classical approximation is not valid, even though $m_X v_0/ m_\phi \gg 1$.}  The crosses show the example parameters from Fig.~\ref{sigmaVv} for the resonant ($m_X = 100$ GeV), Born ($m_X = 4$ GeV), and classical ($m_X = 2$ TeV) regimes.

Most of these resonant features correspond to $s$-wave resonances, and their location in parameter space is given analytically by 
$m_X \approx \pi^2 n^2 m_\phi/(6 \alpha_X)$, where $n=1,2,3$, {\it etc}.  This condition was derived for Sommerfeld enhancements in annihilation~\cite{Cassel:2009wt}, but the same bound state formation arises in scattering as well.  
Taking $\ax\simeq4\times10^{-5}(\mx/{\rm GeV})$ to fix the relic density, we obtain $m_X \approx 6.4 \, {\rm GeV} (m_\phi /{\rm MeV})^{1/2} n$.  This condition matches the locations of resonances in our numerical results.

For asymmetric DM (Fig.~\ref{ResonantPlots}, {\it right}), we take a repulsive-only cross section, $\sigma_T = \sigma_T^{\rm rep}$, and no resonances occur.  We fix $\alpha_X = 10^{-2}$, which provides sufficient depletion of the symmetric $X,\bar X$ density for $m_X \lesssim 300$ GeV (dotted line); above this line, ADM freeze-out would require an additional annihilation channel or a larger $\alpha_X$ (which changes the $\langle \sigma_T \rangle$ contours).  Numerical and analytic results for $\langle \sigma_T \rangle/m_X$ are indicated as in the symmetric case.  

From sub-GeV to multi-TeV DM mass, our results show that a dark force can successfully explain both DM structure and the DM relic density, for $m_\phi \sim 100\, {\rm keV} - {\rm GeV}$.  The $m_X \lesssim$ GeV region corresponds to the Born limit; here, contours at different $v_0$ converge, indicating that $\sigma_T$ is approximately constant in $v$.  At larger $m_X$, $\sigma_T$ is more suppressed at larger $v_0$.  Therefore, possible evidence for DM self-interactions on cluster scales~\cite{Newman:2009qm} may point toward light DM.

{\em IV. Conclusions:} Dark forces may play an important role in the dynamics of DM, analogous to electromagnetic or nuclear forces in the visible sector.  We have shown that a simple, generic model with a dark force can simultaneously explain the DM relic abundance during freeze-out and solve small scale structure anomalies in dwarf galaxies and subhalos, while satisfying constraints on larger galaxy and cluster scales.  We have presented a comprehensive picture of the parameter space of our model, considering both symmetric and asymmetric DM, with attractive or repulsive dark forces.  Within the full parameter space spanning these different cases, we have shown that the DM relic density and self-scattering can accommodate a wide range of DM and mediator masses.  
Importantly for narrowing this range, future astrophysical data favoring or more strongly excluding self-interactions on larger scales would prefer $m_X \lesssim$ GeV or $m_X \gtrsim$ GeV, respectively.  However, N-body simulations over a larger parameter region, including within the resonant regime, would be necessary for detailed comparison with observations.
Experimental tests may also detect the dark force directly, depending on its coupling to visible matter~\cite{Abrahamyan:2011gv}.

{\em Acknowledgments:} We thank the Aspen Center for Physics and the Institute for Advanced Study where part of this work was done. The work of ST, HBY and KZ is supported by the DoE under contract de-sc0007859. The work of HBY and KZ is also supported by NASA Astrophysics Theory Grant NNX11AI17G and by NSF CAREER award PHY 1049896.

{\em Appendix:} We collect analytic formulae for $\sigma_T$ used in the literature.  In the Born regime ($\alpha_X m_X/m_\phi \ll 1$), for both attractive and repulsive forces, a perturbative calculation gives~\cite{Feng:2009hw}
\beq
\sigma_T^{\rm Born} = \frac{8\pi \alpha_X^2}{m_X^2 v^4} \left[ \log(1+R^2) - R^2/(1+R^2) \right] \label{born}
\eeq 
where $R \equiv m_X v/m_\phi$.  In the classical regime ($m_X v/m_\phi \gg 1$), a solution to the classical equations of motion gives for an attractive potential~\cite{Feng:2009hw,Khrapak:2003}
\beq
\sigma_T^{\rm clas} \approx
\left\{\begin{array}{lc}
\frac{4 \pi}{m_\phi^2} \beta^2 \ln\left(1+\beta^{-1}\right) & \beta \lesssim 10^{-1} \\
\frac{8 \pi}{m_\phi^2} \beta^2 / \left(1+1.5 \beta^{1.65}\right) & \; 10^{-1} \lesssim \beta \lesssim 10^3 \\
\frac{\pi}{m_\phi^2} \left(\ln \beta+1-\frac{1}{2} \ln^{-1}\beta \right)^2 & \beta \gtrsim 10^3
\end{array} \right. \label{plasma}
\eeq
where $\beta \equiv 2 \alpha_X m_\phi / (m_X v^2)$, and for the repulsive case~\cite{Khrapak:2004}
\beq
\sigma_T^{\rm clas} \approx
\left\{\begin{array}{lc}
\frac{2 \pi}{m_\phi^2} \beta^2 \ln\left(1+\beta^{-2}\right) & \beta \lesssim 1 \\
\frac{\pi}{m_\phi^2} \big( \ln 2 \beta - \ln \ln 2 \beta  \big)^2& \beta \gtrsim 1
\end{array} \right. \label{plasmaRep} .
\eeq

\end{document}